\documentclass[1p]{elsarticle}
\pdfoutput=1
\usepackage[unicode=true, bookmarks=false,bookmarksnumbered=true, bookmarksopen=true, breaklinks=false, backref=false, pagebackref=false, colorlinks=true]{hyperref}
\hypersetup{pdftitle={SU(2) Lattice Gauge Theory Simulations on Fermi GPUs}, pdfauthor={Nuno Cardoso and Pedro Bicudo}, linkcolor=black, citecolor=black, urlcolor=black, filecolor=black, pdfpagelayout=OneColumn, pdfnewwindow=true, pdfstartview=XYZ, plainpages=false}
\usepackage{color}
\usepackage{amsmath}
\usepackage{amssymb}
\usepackage{braket}
\usepackage[T1]{fontenc}
\usepackage{url}
\usepackage[english]{babel}
\usepackage{dcolumn}
\usepackage{bm}
\usepackage{subfig}
\usepackage{float}
\usepackage{url}
\usepackage{esint}
\usepackage{dsfont}

\newcommand{\Tr}{\mbox{\rm Tr}}
\newcommand{\ReC}{\mbox{\rm Re}}
\newcommand\T{\rule{0pt}{2.6ex}}
\newcommand\B{\rule[-1.2ex]{0pt}{0pt}}

\usepackage{verbatim}

\journal{Journal of Computational Physics}

\begin{document}
\begin{frontmatter}

\title{SU(2) Lattice Gauge Theory Simulations on Fermi GPUs}
\author{Nuno Cardoso\corref{cor1}}
\ead{nunocardoso@cftp.ist.utl.pt}
\author{Pedro Bicudo}
\ead{bicudo@ist.utl.pt}
\address{CFTP, Departamento de F\'{\i}sica, Instituto Superior T\'{e}cnico, Av. Rovisco Pais, 1049-001 Lisboa, Portugal}

\cortext[cor1]{Corresponding author}

\begin{abstract}
In this work we explore the performance of CUDA in quenched lattice SU(2) simulations. CUDA, NVIDIA Compute Unified Device Architecture, is a hardware and software architecture developed by NVIDIA for computing on the GPU.
We present an analysis and performance comparison between the GPU and CPU in single and double precision.
Analyses with multiple GPUs and two different architectures (G200 and Fermi architectures) are also presented.
In order to obtain a high performance, the code must be optimized for the GPU  architecture, i.e., an  implementation that exploits the memory hierarchy of the CUDA programming model.

We produce codes for the Monte Carlo generation of SU(2) lattice gauge configurations,
for the mean plaquette, for the Polyakov Loop at finite T and for the Wilson loop.
We also present results for the potential using many configurations ($50\ 000$) without smearing and almost $2\ 000$ configurations with APE smearing.
With two Fermi GPUs we have achieved an excellent performance of $200 \times$ the speed over one CPU, in single precision, around 110 Gflops/s.
We also find that, using the Fermi architecture, double precision computations for the static quark-antiquark potential are not much slower (less than $2 \times$ slower) than single precision computations.
\end{abstract}

\begin{keyword}
CUDA \sep GPU \sep Fermi \sep SU(2) Lattice Gauge Theory
\MSC 12.38.Gc \sep 07.05.Bx \sep 12.38.Mh \sep 14.40.Pq
\end{keyword}

\end{frontmatter}

\section{Introduction}
Graphics Processing Units (GPUs) have become important in providing processing power for high performance computing applications.
CUDA \cite{Kirk2010,NVIDIA:Manual} is a proprietary API and set of language extensions that works only on NVIDIA's GPUs and call a piece of code that runs on the GPU, a kernel.

In 2007, NVIDIA released CUDA for GPU computing as a language extension to C. CUDA makes the GPU programming and computing development easier and more efficient than the earlier attempts, using OpenGL and associated shader languages, in which it was necessary to translate the computation to a graphics language \cite{Egri:2006zm}.

The most successful theories that describe elementary particle physics are the so called gauge theories. SU(2) is an interesting gauge group, either to simulate the electroweak theory, or to use as a simplified case of the SU(3) gauge group of the strong interaction.
Gauge theories can be addressed by lattice field theory in a non-perturbative approximation scheme based on the path integral formalism in which space-time is discretized.
Quantities in the form of a path integral can be transformed to Euclidean space-time, which can be evaluated numerically using Monte Carlo simulations allowing us to use statistical mechanics methods.

Generating SU(N) lattice configurations is a highly demanding task computationally and requires advanced computer architectures such as CPU clusters or GPUs. Compared with CPU clusters, GPUs are easier to access and maintain, as they can run on a local desktop computer.
There are some groups \cite{Clark:2009wm,Chiu:2011rc,Hayakawa:2010gm} using GPUs to accelerate their lattice simulations, however this is for the full Lagrangian description. 

In this work, we make use of the new GPU technologies to accelerate the calculations in pure gauge lattice SU(2). Note that pure gauge lattice simulations do not include the full Lagrangian description, i.e., dynamical fermions. In particular, we are able to perform our computations integrally in the GPU, thus reaching a quite higher benchmarks when compared with previous computations partially done in the GPU and partially done in the CPU.

This paper is divided in 6 sections. In section 2, we present a brief description on how to generate lattice SU(2) configurations and  in section 3 we give an overview of the GPU hardware and the CUDA programming model. In section 4 we show how to generate lattice SU(2) configurations and calculate the static quark-antiquark potential in one GPU or in multiple GPUs. In section 5 we present the GPU performance over one CPU core, as well as results for the mean average plaquette and Polyakov loop for different $\beta$ and lattice sizes. We also present the static quark-antiquark potential with and without APE smearing. Finally, in section 6, we conclude.

\section{SU(2) Lattice Gauge Theory}
In this section, we describe the heat bath
algorithm for generating SU(2) configurations
\cite{Creutz:1980zj,Creutz:1980zw}.
In SU(2), any group element $U$ may be parametrized in the form,
\begin{equation}
U=a_0\, \mathds{1} + i\mathbf{a}\cdot \sigma \ ,
\end{equation}
where the $\mathbf{\sigma}$ are the usual Pauli matrices and where
\begin{equation}
a^2=a_0^2+\mathbf{a}^2=1 \ .
\end{equation}
This condition defines the unitary hyper-sphere surface $S^3$ and
\begin{equation}
\Tr\, U=2a_0,\quad \quad U U^{\dagger}=U^{\dagger}U=\mathds{1},\quad \quad \det U = 1 \ .
\end{equation}
The invariant Haar group measure is given by
\begin{equation}
dU=\frac{1}{2\pi^2}\delta\left(a^2-1\right)d^4a \ ,
\end{equation}
where $1/(2\pi^2)$ is a normalization factor.

\begin{figure}
\begin{centering}
    \includegraphics[width=10cm]{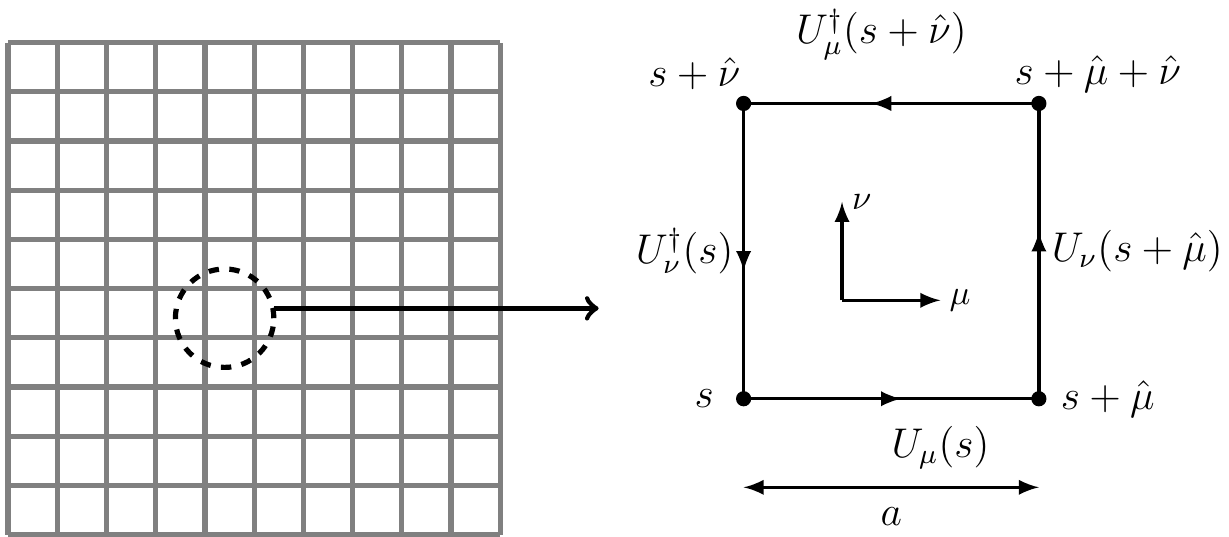}
\par\end{centering}
\caption{Plaquette $P_{\mu\nu}(s)$.}
\label{fig:plaq}
\end{figure}

In order to update a particular link, we need only to consider the contribution to the action from the six plaquettes containing that link, the staple $V$. The plaquette is illustrated in Fig. \ref{fig:plaq}.
Notice that the pure gauge SU(2) lattice action is composed by the sum of all possible plaquettes,
but all the other plaquettes factor out from the expectation value of a particular link.
The distribution to be generated for every single link is given by
\begin{equation}
dP(U)\propto \exp \left[\frac{1}{2} \beta \Tr(UV)\right]\ ,
\end{equation}
where $\beta=4/g_0^2$, $g_0$ is the coupling constant. We apply a useful property of SU(2) elements, that any sum of them is proportional to another SU(2) element $\tilde{U}$,
\begin{equation}
\tilde{U}=\frac{V}{\sqrt{\det V}}=\frac{V}{k}\ .
\end{equation}
Using the invariance of the group measure, we obtain
\begin{equation}
dP\left(U\tilde{U}^{-1}\right) \propto \exp\left[\frac{1}{2}\beta k \Tr U\right]dU = \exp\left[\beta k a_0\right] \frac{1}{2\pi^2}\delta\left(a^2-1\right)d^4 a\ .
\end{equation}
Thus, we need to generate $a_0\in [-1,1]$ with distribution,
\begin{equation}
P\left(a_0\right) \propto \sqrt{1-a_0^2} \exp\left(\beta k a_0\right)\ .
\end{equation}
The components of $\mathbf{a}$ are generated randomly on the 3D unit sphere in a four dimensional space with exponential weighting along the $a_0$ direction.
Once the $a_0$ and $\mathbf{a}$ are obtained in this way, the new link is updated,
\begin{equation}
U'=U\tilde{U}^{-1}\ .
\end{equation}

In order to accelerate the decorrelation of subsequent lattice  configurations, we can employ the over-relaxation algorithm,
\begin{equation}
U_\text{new} = \frac{\Sigma^\dagger}{\left|\Sigma\right|} U^\dagger\frac{\Sigma^\dagger}{\left|\Sigma\right|}\ ,
\end{equation}
where $\Sigma$ is the staple,
\begin{equation}
	\Sigma = \sum_{\mu\neq\nu}(U_{x,\nu}U_{x+\hat{\nu},\mu}U^\dagger_{x+\hat{\mu},\nu} + U^\dagger_{x-\hat{\nu},\nu}U_{x-\hat{\nu},\mu}U_{x-\hat{\nu}+\hat{\mu},\nu})\ ,	
\end{equation}
and $|\Sigma| = \sqrt{\det\Sigma}$.

The simplest measurement that can be done in the lattice is the average plaquette. The average plaquette, $\Braket{P}$, is given by,
\begin{equation}
\Braket{P} = \frac{1}{V}\sum_{s\,\in\, \text{lattice}} \sum_{{\mu,\nu\atop  \mu<\nu}} P_{\mu\nu}(s)\ ,
\label{eq:plaq}
\end{equation}
where $V$ is the lattice volume and $P_{\mu\nu}(s)$, see Fig. \ref{fig:plaq}, is
\begin{equation}
P_{\mu\nu}(s) = 1 - \frac{1}{2} \ReC \Tr \left[ U_\mu(s)U_\nu(s+\hat{\mu})U_\mu^\dagger(s+\hat{\nu})U_\nu^\dagger(s) \right] \ .
\end{equation}

\begin{figure}
\begin{centering}
    \includegraphics[width=5cm]{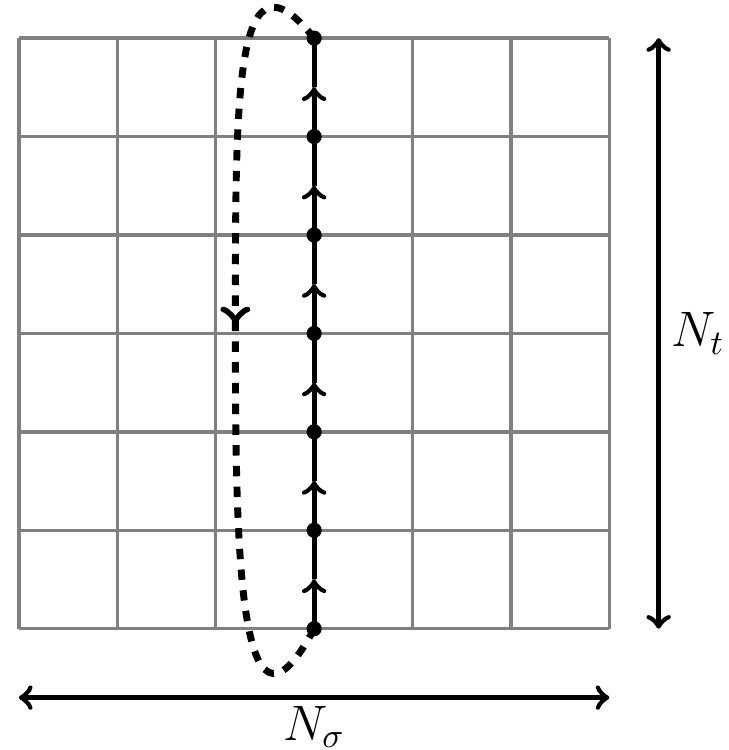}
\par\end{centering}
\caption{Polyakov loop.}
\label{fig:ploop}
\end{figure}

Another interesting operator that can be calculated in the lattice is the expectation value of the Polyakov loop, $\Braket{L}$ \cite{McLerran:1980pk,Engels1988289},
\begin{equation}
\Braket{L} = \frac{1}{N_\sigma} \sum_{\mathbf{x}} L\left(\mathbf{x}\right) \ ,
\end{equation}
where $N_\sigma=N_x\times N_y\times N_z$.
The product of link variables on the temporal direction
$L\left(\vec{x}\right)$  is  depicted in Fig. \ref{fig:ploop}, some times this is called a Wilson line,
\begin{equation}
L\left(\mathbf{x}\right) = \frac{1}{2} \Tr \prod_{t=0}^{N_t-1} U_4(\mathbf{x},t) \ ,
\end{equation}
where $U_4$ is the link along the temporal direction. Since we employ periodic boundary conditions in time direction, $U_\mu(\mathbf{x},0)=U_\mu(\mathbf{x},N_t)$, and in space direction, this is equivalent to a closed loop. The expectation value of the Polyakov loop is the order parameter for the deconfinement transition on an infinite lattice \cite{McLerran:1980pk}.
The order parameter measures the free energy, $F_q$ of a single static (infinite mass) quark at temperature $T$,
\begin{equation}
\Braket{L}\propto \exp\left(- \frac{F_q}{T} \right) \ ,
\end{equation}
where $T$ is connected to the lattice spacing $a$ by
\begin{equation}
T=\frac{1}{N_t \,a} \ .
\end{equation}
When $\Braket{L}=0$, the free energy of the quark increases arbitrarily with the volume, and this is interpreted as a signal of quark confinement. When $\Braket{L}\neq 0$, the free energy of the quark tend to a constant for large volume, and this is interpreted as a signal of deconfinement.

\begin{figure}
\begin{centering}
    \includegraphics[width=5cm]{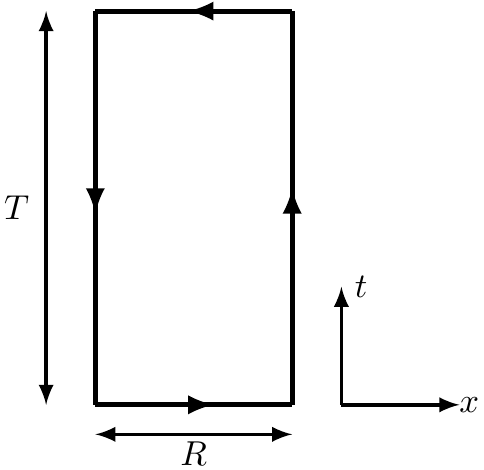}
\par\end{centering}
    \caption{Wilson Loop operator on the lattice.}
    \label{WL}
\end{figure}

We can also extend the square of size $1\times 1$, i.e., the plaquette, to  construct an operator with a larger size, the Wilson loop. The Wilson loop, depicted in Fig. \ref{WL}, is given by,
\begin{eqnarray}
W(R, T) & = & \Tr\left[ U_\mu (0,0) \cdots U_\mu((R-1)\hat{\mu},0)U_4(R\hat{\mu},0)\cdots U_4(R\hat{\mu},T-1) \right. \nonumber \\
  &  & \left. U_\mu^\dagger((R-1)\hat{\mu},T)\cdots U_\mu^\dagger(0,T)U_4^\dagger(0,T-1)\cdots U_4^\dagger(0,0) \right] \ ,
 \end{eqnarray}
where $R$ is the spatial direction and $T$ is the temporal direction.
Note that the smallest non-trivial Wilson loop on the lattice is the plaquette.
The mean value of the Wilson loop is utilized to compute the static quark-antiquark potential.

In order to improve the signal to noise ratio of the Wilson loop, we can use the APE smearing.
The APE smearing is a gauge equivariant prescription for averaging a link $U_{\mu}(x)$ with its nearest neighbours,
\begin{eqnarray}
    U_{\mu}\left(s\right) & \rightarrow & P_{SU(2)}\frac{1}{1+6w}\Big(U_{\mu}\left(s\right) + w \sum_{\mu\neq\nu}U_{\nu}\left(s\right) U_{\mu}\left(s+\hat{\nu}\right)U_{\nu}^{\dagger}\left(s+\hat{\mu}\right)\Big)\ ,
\end{eqnarray}
where $P_{SU(2)}$ is a projector back onto the SU(2) group, $w = 0.2$ and iterate this procedure 25 times in the spatial direction.
Empirically, it is seen that using a smeared operator helps to improve ground-state overlap dramatically.

\section{Cuda Programming Model}

CUDA \cite{Kirk2010,NVIDIA:Manual} is the hardware and software that enables NVIDIA GPUs to execute programs written with languages such as C, C++, Fortran, OpenCL and DirectCompute.

CUDA programs call parallel kernels, each of which executes in parallel across a set of parallel threads.
These threads are then organized, by the compiler or the programmer, in thread blocks and grids of thread blocks.

The GPU instantiates a kernel program on a grid of parallel thread blocks.
Within the thread blocks, an instance of the kernel will be executed by each thread, which has a thread ID within its thread block, program counter, registers, per-thread private memory, inputs, and output results.
Thread blocks are sets of concurrently executing threads, cooperating among themselves by barrier synchronization and shared memory. Thread blocks also have block ID's within their grids.
A grid is an array of thread blocks. This array executes the same kernel, reads inputs from global memory, writes results to global memory, and synchronizes between dependent kernel calls.

In the CUDA parallel programming model, each thread has a per-thread private memory space used for register spills, function calls, and C automatic array variables.
Each thread block has a per-Block shared memory space used for inter-thread communication, data sharing, and result sharing in parallel algorithms.
Grids of thread blocks share results in Global Memory space after kernel-wide global synchronization.

\begin{figure}
\begin{centering}
    \includegraphics[width=10cm]{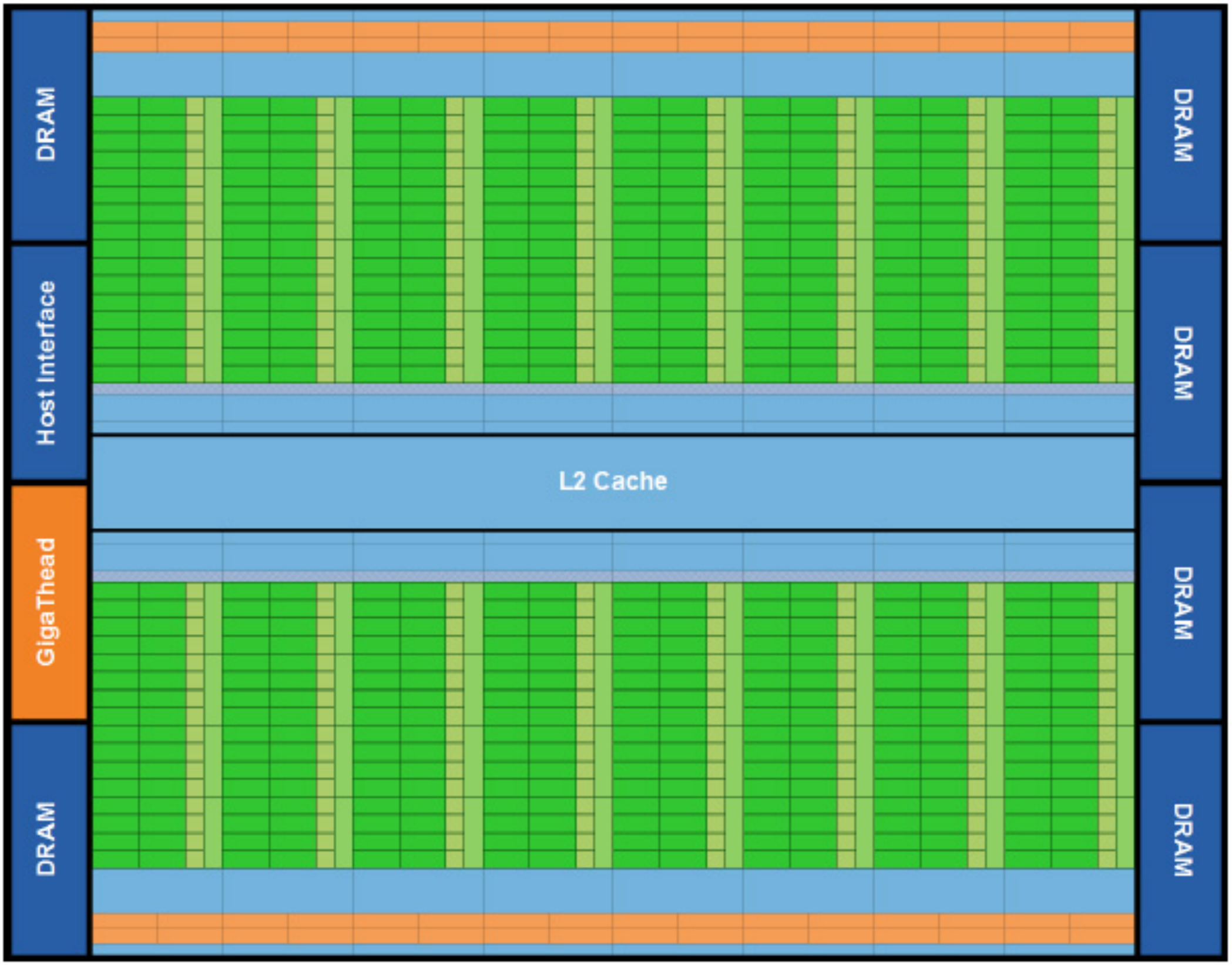}
\par\end{centering}
    \caption{Fermi Architecture. Fermi's 16 streaming multiprocessors are positioned around a common L2 cache.
Each streaming multiprocessors is a vertical rectangular strip that contain an orange portion (scheduler and dispatch), a green portion (execution units), and light blue portions (register file and L1 cache) \cite{Fermi2010}.}
    \label{Fermi_Architecture}
\end{figure}

In the hardware execution view, CUDA's hierarchy of threads maps to a hierarchy of processors on the GPU; a GPU executes one or more kernel grids; a streaming multiprocessor (SM) executes one or more thread blocks; and CUDA cores and other execution units in the SM execute threads.
The SM executes threads in groups of 32 threads called a warp.
While programmers can generally ignore warp execution for functional correctness and think of programming one thread, they can greatly improve performance by having threads in a warp executing the same code path and accessing memory in nearby addresses.

The first Fermi based GPU implemented with 3.0 billion transistors, features up to 512 CUDA cores, organized in 16 SMs of 32 cores each.
A CUDA core executes a floating point or integer instruction per clock for a thread. In Fig. \ref{Fermi_Architecture} and Table \ref{nvidia_arch_sum} we present the details of the Fermi architecture.
The GPU has six 64-bit memory partitions, for a 384-bit memory interface, and supports  up to a total of 6 GB of GDDR5 DRAM memory.
The connection of the GPU to the CPU is made by a host interface via PCI-Express.
GigaThread global scheduler distributes thread blocks to SM thread schedulers.

\begin{table}
\begin{centering}
\begin{tabular}{|c|c|c|c|}
\hline
\T\B GPU & G80 & GT200 & Fermi\tabularnewline
\hline
\hline
\T\B Transistors & 681 million & 1.4 billion & 3.0 billion\tabularnewline
\hline
\T\B CUDA cores & 128 & 240 & 512\tabularnewline
\hline
\T\B Double precision & None & 30 FMA & 256 FMA \tabularnewline
\T\B floating point capability &  & ops/clock & ops/clock\tabularnewline
\hline
\T\B Single precision & 128 MAD& 240 MAD& 512 MAD \tabularnewline
\T\B floating point capability & ops/clock & ops/clock & ops/clock \tabularnewline
\hline
\T\B Warp schedulers (per SM) & 1 & 1 & 2 \tabularnewline
\hline
\T\B Special function & 2 & 2 & 4  \tabularnewline
\T\B units (SFUs)/SM &  &  &   \tabularnewline
\hline
\T\B Shared memory & 16KB & 16KB & Configurable \tabularnewline
\T\B (per SM) &  & & 48KB or 16KB \tabularnewline
\hline
\T\B L1 cache & None & None & Configurable \tabularnewline
\T\B (per SM) &  &  & 16KB or 48KB \tabularnewline
\hline
\T\B L2 cache (per SM) & None & None & 768KB \tabularnewline
\hline
\T\B ECC memory support & No & No & Yes \tabularnewline
\hline
\T\B Concurrent kernels & No & No & Up to 16 \tabularnewline
\hline
\T\B Load/Store address width & 32-bit & 32-bit & 64-bit \tabularnewline
\hline
\end{tabular}
\par\end{centering}
\caption{NVIDIA's architecture specifications (SM means Streaming Multiprocessor) Source \cite{Fermi2010}.}
\label{nvidia_arch_sum}
\end{table}

The Fermi architecture \cite{Fermi2010} represents the most important improvement in GPU architecture since the original G80, an early vision on unified graphics and computing parallel processor. GT200 extended its performance and functionality.
Table \ref{nvidia_arch_sum} shows the details between the different architectures (G80, GT200 and Fermi architectures).
With Fermi, NVIDIA used the knowledge from the two prior processors and all the applications that were written for them, and employed a completely new approach to design and to create the world's first computational GPU.

The Fermi team designed a processor, Fig. \ref{Fermi_Architecture}, that highly increases not only raw compute horsepower, but also, at the same time,  the programmability and computational efficiency using architectural innovations.
They made improvements in double precision performance, a true cache hierarchy since some algorithms cannot take advantage of the Shared memory resources (NVIDIA Parallel DataCache hierarchy with configurable L1 and unified L2 caches), have more shared memory, faster context switching and faster atomic operations.

In all GPUs architectures, it is necessary to take into account the following performance considerations: memory coalescing, shared memory bank conflicts, control-flow divergence, occupancy and kernel launch overheads.

\section{Mapping Lattice SU(2) to GPU}
In this section, we discuss the parallelization scheme for generating pure gauge SU(2) lattice configurations.

A CUDA application works by spawning a very large number of threads on the GPU which are executed in parallel.
The threads are grouped in thread blocks and the entire collection of blocks is called a grid.
CUDA provides primitives that allow the synchronization within a thread block. However, it is not possible to synchronize threads within different thread blocks.
In order to avoid the penalty for high latency, we must ensure a high multiprocessor occupancy, i.e., each multiprocessor should have many threads simultaneously loaded and waiting for execution.
In this work, we assign one thread to each lattice site and in all runs we maintain the thread block size fixed.
Since CUDA only supports thread blocks up to 3D and grids up to 2D, and the lattice needs four indexes, we use 3D thread blocks, one for $t$, one for $z$ and one for both $x$ and $y$. We then reconstruct the other index inside the kernel.

We place most of the constants needed by the GPU, like the number of points in the lattice, in the constant memory using \mbox{\texttt{cudaMemcpyToSymbol}}, as in the following example
\begin{verbatim}
cudaMemcpyToSymbol( "Nx", &Nx, sizeof(int) );
\end{verbatim}
The code to obtain the four indices of the 4D hypercube, when using a single GPU, inside the kernel is
\begin{verbatim}
int blockIdxz = __float2int_rd(blockIdx.y * invblocky);
int blockIdxy = blockIdx.y - __umul24(blockIdxz, blocks_y);
int ij = __mul24(blockIdx.x, blockDim.x) + threadIdx.x;

//Index's of 4D hyper-cube
int i = mod(ij, Nx);
int j = __float2int_rd(ij / Nx);
int k = __mul24(blockIdxy, blockDim.y) + threadIdx.y;
int t = __mul24(blockIdxz, blockDim.z) + threadIdx.z;
\end{verbatim}
and outside the kernel we define,
\begin{verbatim}
threads_x = mineq(Nx * Ny, 16);
threads_y = mineq(Nz, 4);
threads_z = mineq(Nt, 4);

blocks_x = (Nx * Ny + threads_x - 1) / threads_x;
blocks_y = (Nz + threads_y - 1) / threads_y;
blocks_z = (Nt + threads_z - 1) / threads_z;
	
block = make_uint3(threads_x, threads_y, threads_z);
grid = make_uint3(blocks_x, blocks_y * blocks_z, 1);
invblocky = 1.0f / (T)blocks_y;
\end{verbatim}
where \mbox{\texttt{mineq()}} is a function that returns the minimum value. A kernel is then defined, for example, as
\begin{verbatim}
Cold_Start<T4><<< grid, block >>>(lattice_d);
\end{verbatim}

Note that in the Polyakov loop kernel we only need three indexes and we can use the 3D thread blocks, i.e., in the kernel, we use
\begin{verbatim}
int blockIdxz = __float2int_rd(blockIdx.y * invblocky_3D);
int blockIdxy = blockIdx.y - __umul24(blockIdxz,blocky_3D);
int i     = __mul24(blockIdx.x,blockDim.x) + threadIdx.x;
int j     = __mul24(blockIdxy ,blockDim.y) + threadIdx.y;
int k     = __mul24(blockIdxz ,blockDim.z) + threadIdx.z;
\end{verbatim}
and each thread make the temporal link multiplication from $t=0$ to $t=N_t-1$ and the number of thread blocks and the number of block is defined as,
\begin{verbatim}
threads_x = mineq(Nx, 8);
threads_y = mineq(Ny, 8);
threads_z = mineq(Nz, 8);

blocks_x = (Nx + threads_x - 1) / threads_x;
blocks_y = (Ny + threads_y - 1) / threads_y;
blocks_z = (Nz + threads_z - 1) / threads_z;

block_3D = make_uint3(threads_x, threads_y, threads_z);
grid_3D = make_uint3(blocks_x, blocks_y * blocks_z, 1);
invblocky_3D = 1.0f/(T)blocks_y;
blocky_3D = blocks_y;
\end{verbatim}

Since memory transfers between CPU and GPU are very slow comparing with other GPU memory and in order to maximize the GPU performance, we should only use this feature when it is extremely necessary.
Hence, we only use CPU/GPU memory transfers in three cases: in the initial array of seeds for the random number generator in the GPU, in the end of the kernel to perform the sum over all lattice sites (copy the final result to CPU memory) and when using multi-GPUs (exchange the border cells between GPUs).

The kernels developed for this work are:
\begin{itemize}
	\item Random number generator, RNG;
	\item Lattice initialization:
	\begin{itemize}
		\item Cold start, $U=\mathds{1}$;
		\item Hot start, random SU(2) matrix;
		\item Read a configuration from input file.
	\end{itemize}
	\item Heat bath algorithm;
	\item Over-relaxation method;
	\item Plaquette (for each site);
	\item Polyakov Loop (for each site);
	\item Wilson Loop (for each site);
	\item APE Smearing;
	\item Parallel reduction. Sum over all sites of an array. This kernel performs a sum over all sites after calculation of the plaquette, Polyakov loop and Wilson loop.
\end{itemize}

For the generation of the random numbers needed in the hot start lattice initialization and in the heat bath algorithm, we use a linear congruential random number generator (LCRNG) \cite{press}, given by
\begin{equation}
x_{i+1,j}=(a\,x_{i,j} + b)\mod m\ ,
\end{equation}
and
\begin{equation}
x_{0,j+1}=(c\,x_{0,j})\mod m\ ,
\end{equation}
with $a=1664525$, $b=1013904223$, $c=16807$, $m=2147483647$ and $x_{0,0}=1$. We generate the first random numbers $x_{0,j}$ in the CPU and then copy the array to the GPU. Therefore, we can generate a different random number in each GPU thread.

The LCRNG is used only in the performance tests since this type of random number generator is not suitable for production running. However, in the results we use the random number generator included with the NVIDIA Toolkit 3.2 RC2, CURAND library \cite{NVIDIA:CURAND}.

For the lattice array we cannot use in CUDA a four dimensional array to store the lattice. Therefore we use a 1D array with size $N_x\times N_y\times N_z\times N_t\times Dim$ and a \mbox{\texttt{float4}}, in the case of single precision, or \mbox{\texttt{double4}}, for double precision, to store the generators of SU(2) ($a_0$, $a_1$, $a_2$ and $a_3$). Then, we need to construct all the CUDA operators to make all the operations needed. In this way, we only need four floating point numbers per link instead of having a $2\times 2$ complex matrix.
In order to select single or double precision, we use templates in the code.

In the heat bath and over-relaxation methods, since we need to calculate the staple at each link direction and given the GPU architecture, we use the chessboard method, calculating the new links separately by direction and by even and odd sites.

The Plaquette, Polyakov Loop and Wilson Loop kernels are used to calculate the plaquette, the Polyakov loop and the Wilson loop by lattice site. In the end we need to perform the sum over all lattice sites. To make this sum, we use the parallel reduction code (kernel 6) in the NVIDIA GPU Computing SDK package \cite{NVIDIA_CUDA_SOFT,parallelreduction}.

\begin{figure}
\begin{centering}
    \subfloat[\label{Grid}]{
\begin{centering}
    \includegraphics[width=4.0cm]{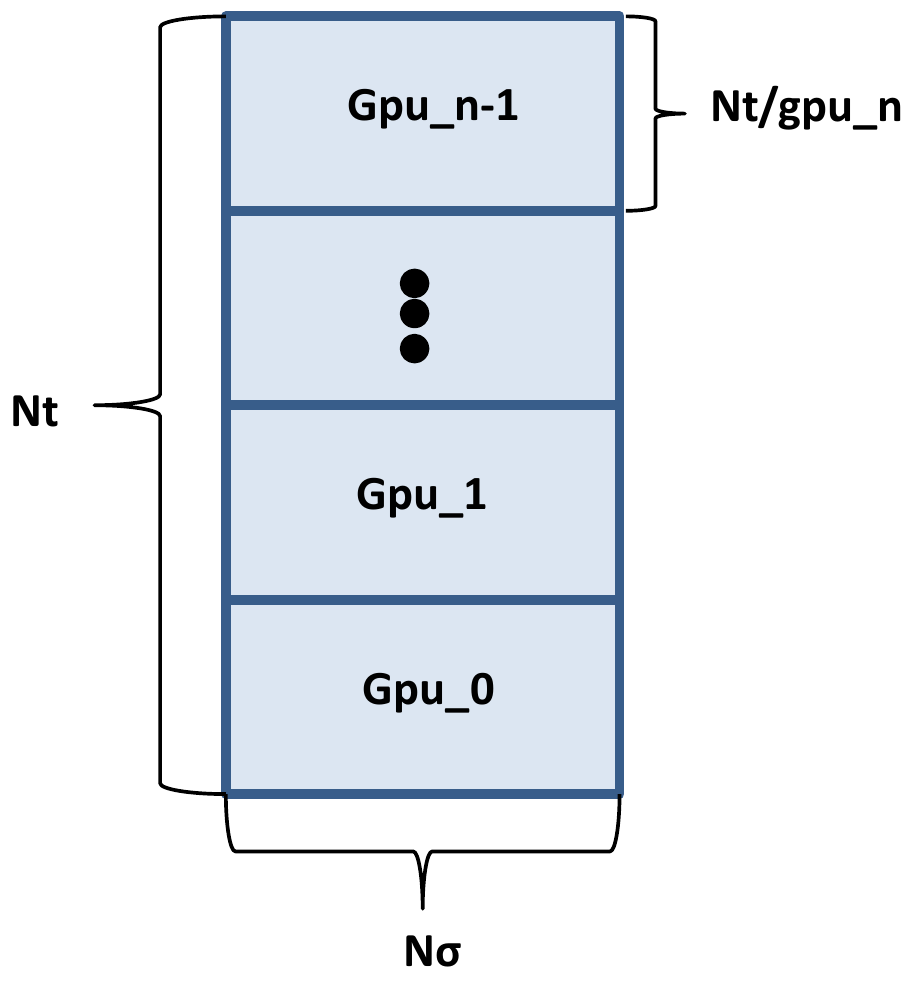}
\par\end{centering}}
    \subfloat[\label{Grid2}]{
\begin{centering}
    \includegraphics[width=4cm]{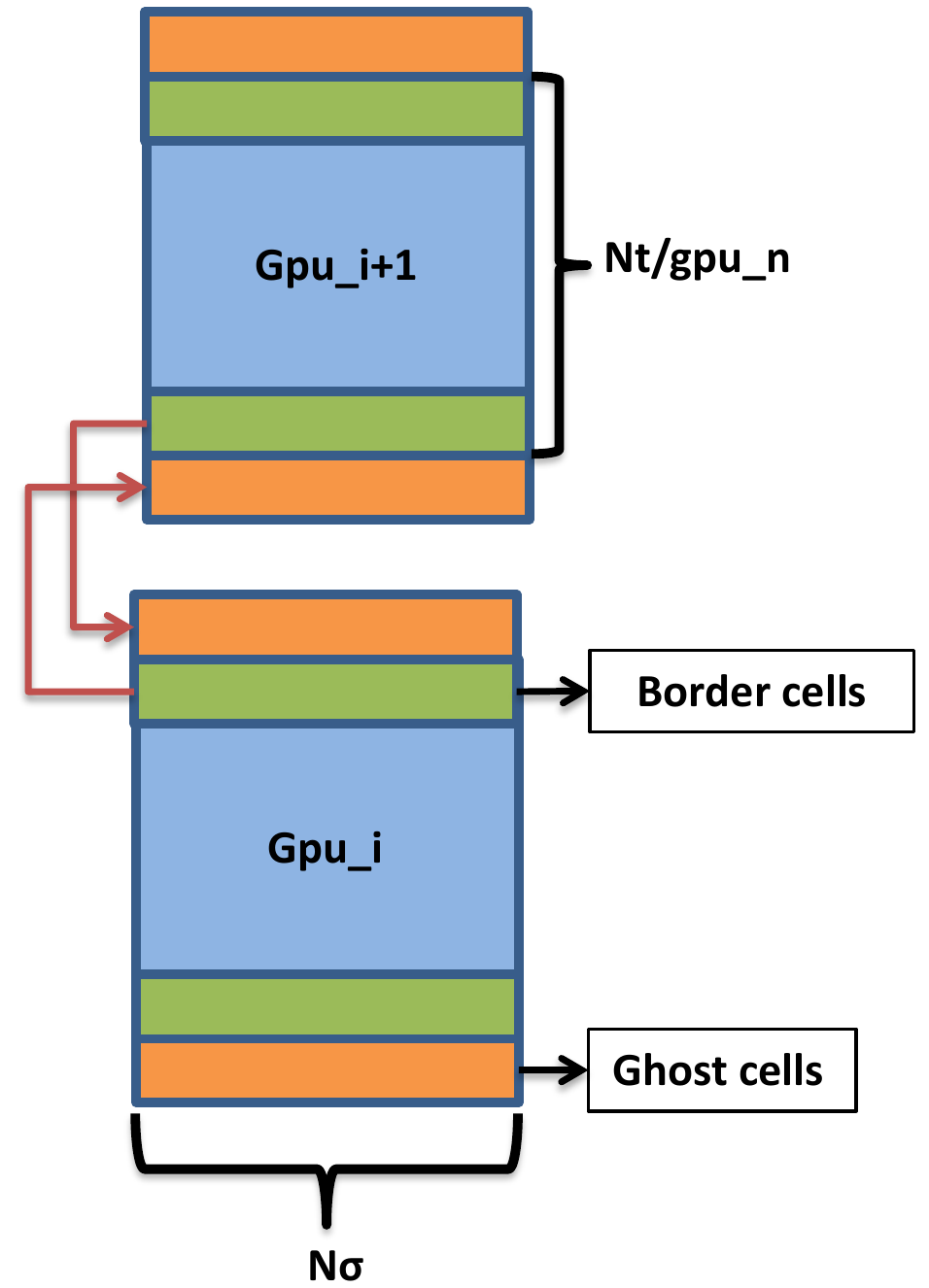}
\par\end{centering}}
\par\end{centering}
    \caption{Schematic view of the lattice array handled by each GPU.}
    \label{openmp_grid}
\end{figure}

Although CUDA neither supports explicitly double textures nor supports double4 textures, it is possible to bind a double4 array to a texture and then retrieve double4 values. This can be done by declaring the texture as \mbox{\texttt{int4}} and then using \mbox{\texttt{\_\_hiloint2double}} to cast it to double, as in the following code example:
\begin{verbatim}
texture<int4, 1, cudaReadModeElementType> tex_lattice_double;
\end{verbatim}
\begin{verbatim}
__device__ double4 fetch_lat(double4 *x, int i){
#if __CUDA_ARCH__ >= 130
    // double requires Compute Capability 1.3 or greater
    if (UseTex)
    {
          int4 v = tex1Dfetch(tex_lattice_double, 2 * i);
          int4 u = tex1Dfetch(tex_lattice_double, 2 * i + 1);
          return make_double4(__hiloint2double(v.y, v.x),
                              __hiloint2double(v.w, v.z),
                              __hiloint2double(u.y, u.x),
                              __hiloint2double(u.w, u.z));
    }
    else
        return x[i];
#else
    return x[i];
#endif
}
\end{verbatim}
moreover, float textures are declared and accessed as,
\begin{verbatim}
texture<float4, 1, cudaReadModeElementType> tex_lattice;
\end{verbatim}
\begin{verbatim}
__device__ float4 fetch_lat(float4 *x, int i){
    if (UseTex)
        return tex1Dfetch(tex_lattice, i);
    else
        return x[i];
}
\end{verbatim}

We now address the multi-GPU approach.
The Multi-GPU part was implemented using CUDA and OPENMP, each CPU thread controls one GPU. Each GPU computes $N_\sigma \times \frac{N_t}{num.\, gpus}$.
The total length of the array in each GPU is then $N_\sigma\times (\frac{N_t}{num.\, gpus} + 2)$, see Fig. \ref{openmp_grid}.
At each iteration, the links are calculated separately by even and odd lattice sites and by the direction $\mu$.
Before calculating the next direction, the border cells in each GPU need to be exchanged between each GPU.
On the border of each lattice, at least one of the neighboring sites is located in the memory of another GPU, see Fig. \ref{Grid2}.
For this reason, the links at the borders of each lattice have to be transferred from one GPU to the GPU handling the adjacent lattice.
In order to exchange the border cells between GPUs it is necessary to copy these cells to CPU memory and then synchronize each CPU thread with the command \mbox{\texttt{\#pragma omp barrier}} before updating the GPU memory, ghost cells.

\section{Results}
Here we present the benchmark results using two different GPU architectures (GT200 and Fermi) in generating pure gauge lattice SU(2) configurations. We also compare the performance with two Fermi GPUs working in parallel in the same mother-board, using CUDA and OPENMP.

Results for the mean average plaquette and Polyakov loop are also presented. Finally, the static quark-antiquark potential is calculated on GPUs using single and double precision. We also present results with smeared and unsmeared configurations, as well as the results obtained for the lattice spacing with $\beta=2.8$. In these results, we didn't use any step of over-relaxation.

Our code can be downloaded from the Portuguese Lattice QCD collaboration homepage \cite{ptqcd}.

\subsection{Performance of Monte Carlo Generator}
In this section, we compare the performance between GPU's, see table \ref{nvidia_gpu_specs}, (two different architectures, NVIDIA GTX 295, GT200 architecture, and NVIDIA GTX 480, FERMI architecture) and a CPU (Intel Core i7 CPU 920, 2.67GHz, 8 MB L2 Cache and 12 GB of RAM).
We compare the performance in generating pure gauge lattice SU(2) configurations and measure the mean average plaquette for each iteration with $\beta=6.0$, hot start initialization and 100 iterations in single and double precision.

\begin{table}
\begin{centering}
\begin{tabular}{|c|c|c|}
\hline
\T\B NVIDIA Geforce GTX & 295 & 480\tabularnewline
\hline
\hline
\T\B Number of GPUs & 2 & 1\tabularnewline
\hline
\T\B CUDA Capability & 1.3 & 2.0\tabularnewline
\hline
\T\B Number of cores & 2$\times$240 & 480\tabularnewline
\hline
\T Global memory & 1792 MB GDDR3 & 1536 MB\tabularnewline
\B & (896MB per GPU) & GDDR5\tabularnewline
\hline
\T\B Number of threads per block & 512 & 1024\tabularnewline
\hline
\T\B Registers per block & 16384 & 32768\tabularnewline
\hline
\T\B Shared memory (per SM) & 16KB B & 48KB or 16KB\tabularnewline
\hline
\T\B L1 cache (per SM) & None & 16KB or 48KB \tabularnewline
\hline
\T\B L2 cache (per SM) & None & 768KB \tabularnewline
\hline
\T\B Clock rate & 1.37 GHz & 1.40 GHz\tabularnewline
\hline
\end{tabular}
\par\end{centering}
\caption{NVIDIA's graphics card specifications used in this work.}
\label{nvidia_gpu_specs}
\end{table}

\begin{figure}
\begin{centering}
    \subfloat[Single precision.\label{fig:spec_sp}]{
\begin{centering}
    \includegraphics[width=6cm]{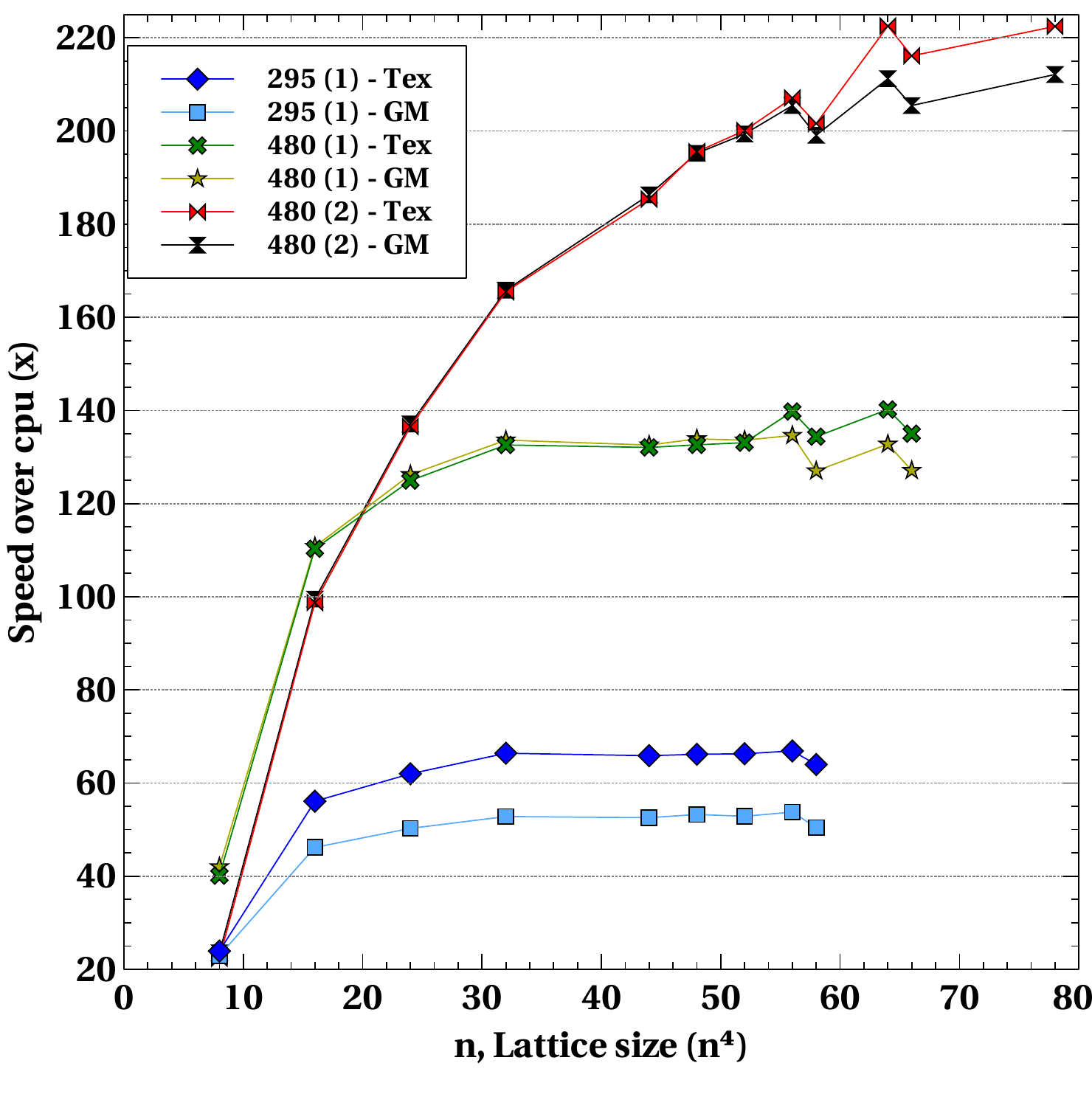}
\par\end{centering}}
    \subfloat[Double precision.\label{fig:spec_dp}]{
\begin{centering}
    \includegraphics[width=6cm]{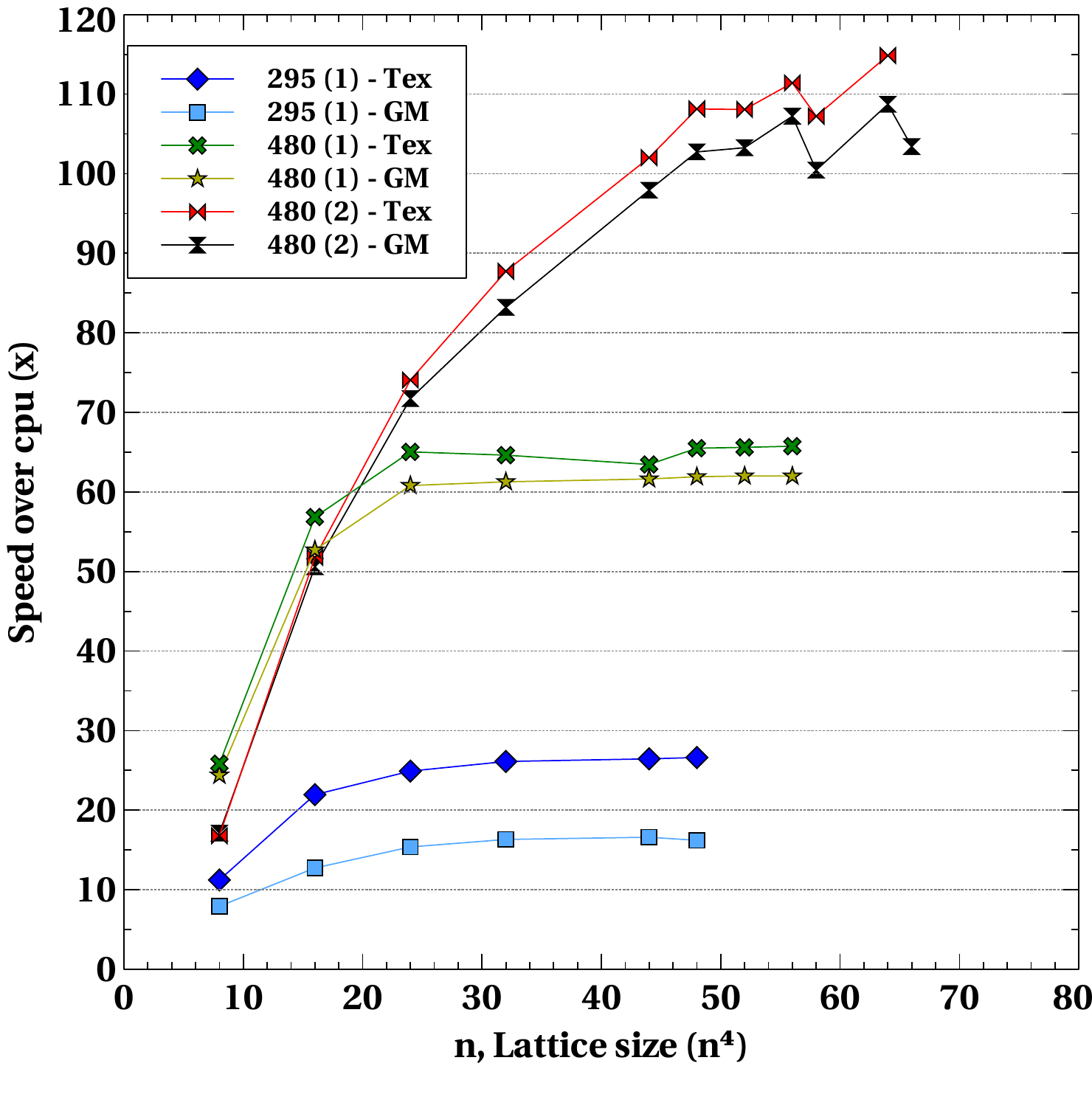}
\par\end{centering}}
\par\end{centering}
    \caption{Performance results. 295 - NVIDIA Geforce 295 GTX; 480 - NVIDIA Geforce 480 GTX; (1) - with 1 GPU; (2) - with 2 GPUs; Tex - using textures; GM - using global memory.}
    \label{fig:spec}
\end{figure}

\begin{figure}
\begin{centering}
    \subfloat[Single precision.\label{fig:flops_sp}]{
\begin{centering}
    \includegraphics[width=6cm]{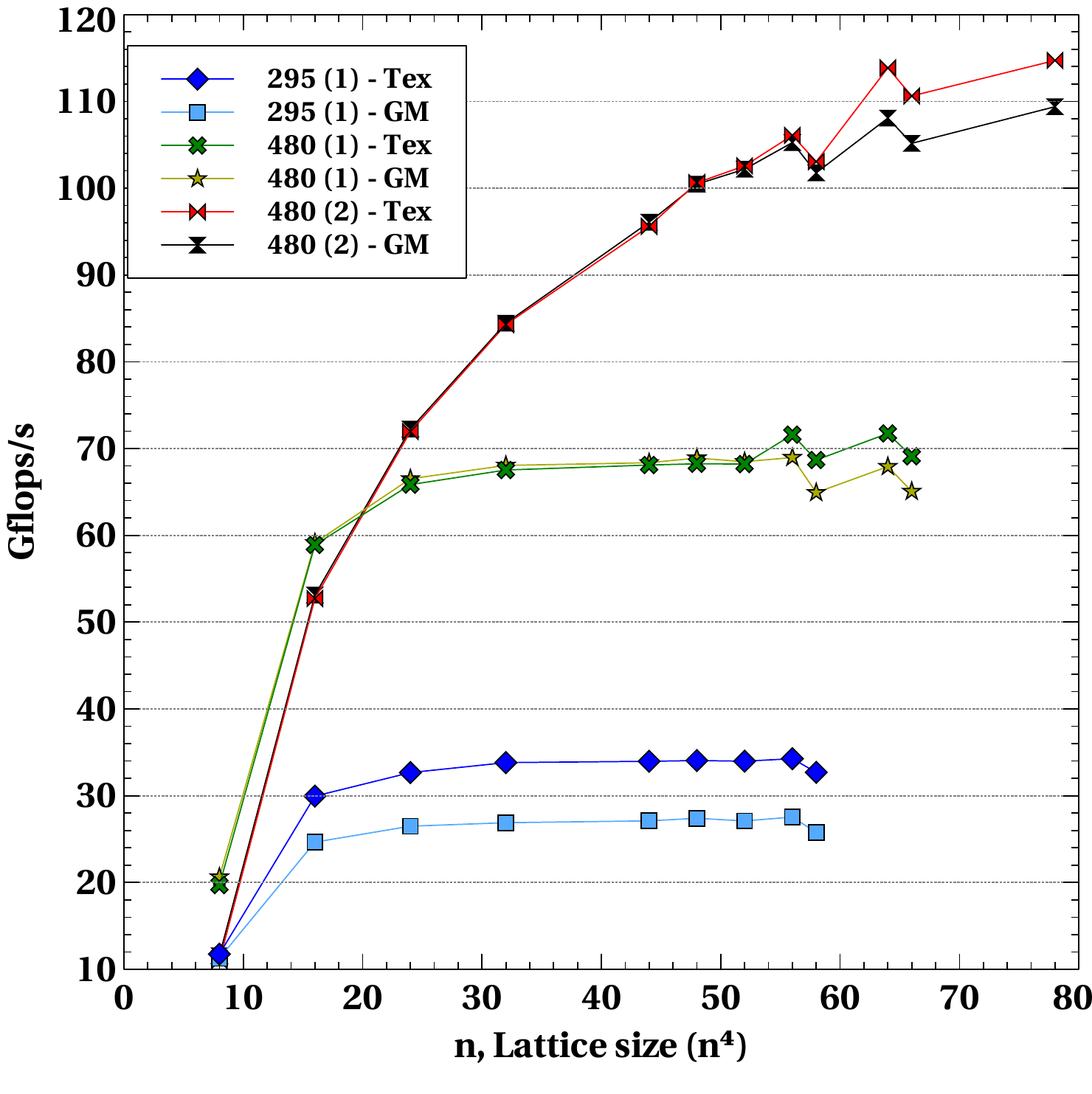}
\par\end{centering}}
    \subfloat[Double precision.\label{fig:flops_dp}]{
\begin{centering}
    \includegraphics[width=6cm]{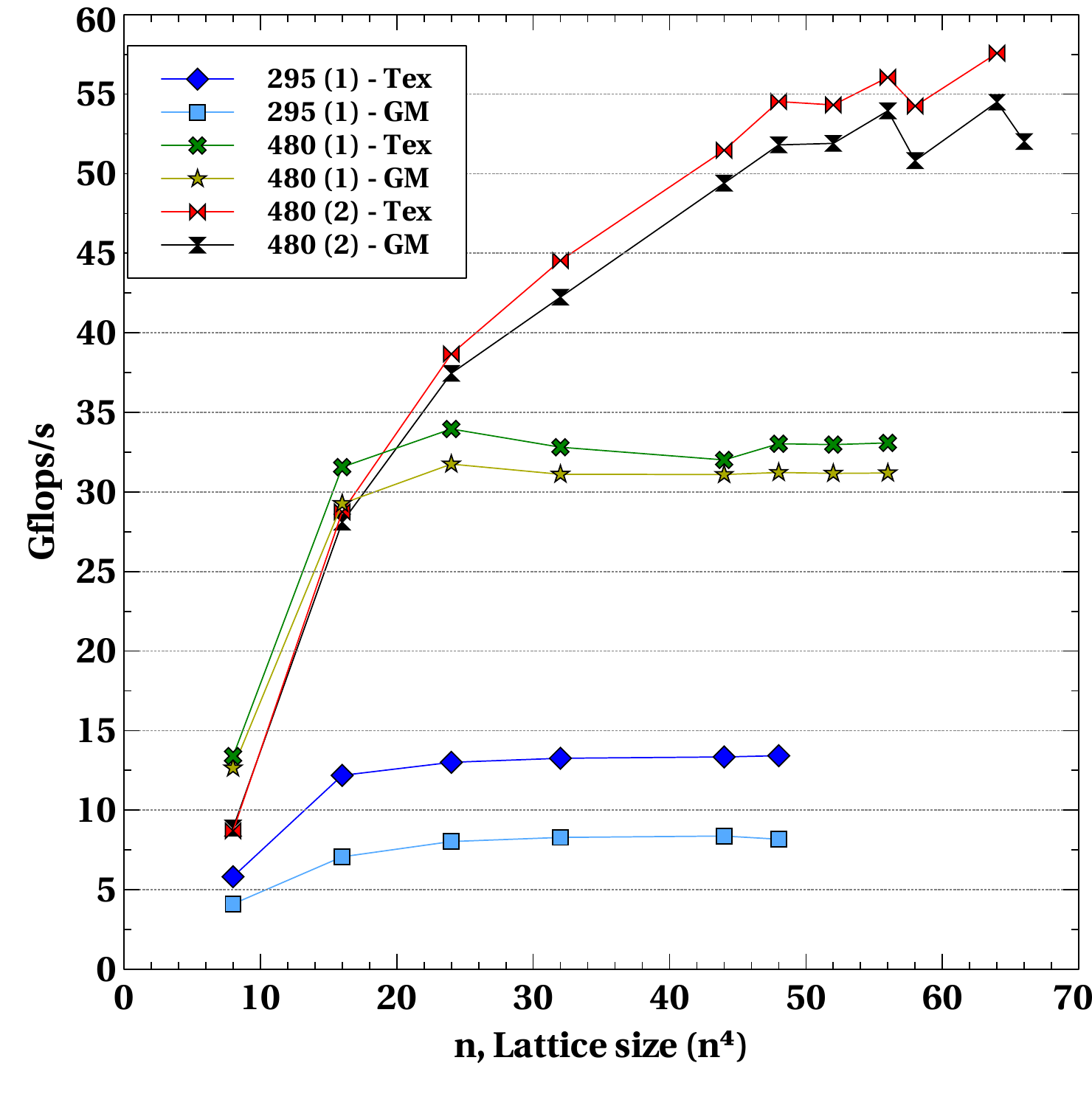}
\par\end{centering}}
\par\end{centering}
    \caption{Performance results in Gflops/s for 100 iterations starting with a random SU(2) matrix and one try to update a gauge link. 295 - NVIDIA Geforce 295 GTX; 480 - NVIDIA Geforce 480 GTX; (1) - with 1 GPU; (2) - with 2 GPUs; Tex - using textures; GM - using global memory. Notice that our code, in one CPU core has a performance of 510 Mflops/s to 520 Mflops/s.}
    \label{fig:spec_flops}
\end{figure}

In Fig. \ref{fig:spec}, we present the performance results using NVIDIA GPUs, NVIDIA GTX 295 (with 2 GPUs per board) and 2 NVIDIA GTX 480 (with 1 GPU per board) versus one CPU core. 
Our CPU code to generate a random SU(2) matrix and the heat bath algorithm is simply the same code we developed for the GPU, except that the memory and process transfers to the GPU are different, as well as the process to sum an array (calculate the mean plaquette value).
Moreover, our CPU code is not implemented with SSE instructions. In Fig. \ref{fig:spec_flops}, we show the GPU performance in Gflops/s. We run the code for 100 iterations, starting with a random SU(2) configuration. In the heat bath algorithm, we only perform one try to update the link. In this way, we measure the flops in all kernels used (kernel to initialize the random SU(2) configuration, heat bath kernel, plaquette kernel and the parallel reduction kernel). Although the GPU peak performance is around one Tflops/s in single performance, the performance achieved by our code, around 70 Gflops/s using one Fermi GPU, is significantly affected by the large memory transfers, i.e., for each try to update one gauge link, we need to copy from global memory 19 links (19$\times$float4(double4)) plus one unsigned int in the random array and to calculate the plaquette at each lattice site we need to copy 24 links (24$\times$float4(double4)). Note that in the heat bath kernel we need to calculate new random numbers but this is not accounted in the number of flops, as well as we only count one instruction for log(), cos(), sin() and sqrt() functions. The CPU (Intel Core i7 CPU 920, 2.67GHz, 8 MB L2 Cache and 12 GB of RAM) performance in one core is almost constant as the lattice size increases, 510-520 Mflops/s.

The memory access inside the GPUs was done using two methods, one using textures and the other one using the global memory in the NVIDIA GTX 295 case and the cache memory in NVIDIA GTX 480. We don't use the shared memory because it is a resource too small to fit in our problem. We only show the performance tests for a maximum lattice array that can fit in our GPU memory. Using only one Fermi GPU, the maximum lattice array size in the GPU memory is $66^4$ and $56^4$ for single and double precision, respectively.

In the Fermi architecture there is not much difference between using textures or accessing to global memory when using single precision. This is because of the new cache hierarchy (L1 and L2 cache). In architectures prior to Fermi, there is no cache hierarchy, therefore, when using textures on these architectures, we can achieve a higher performance in comparison to accessing to the Global memory. However, when using textures there is a limitation of the array size, the maximum width for a 1D texture reference bound to linear memory is $2^{27}$, independent of the GPU architecture.

Splitting the lattice array in four, i.e., one array for each link direction, we can achieve $1.4\times$ the speed over using only one single array to store all the lattice. However, using four arrays makes it harder to add new code, since it forces us to write the code more explicitly and the programming errors are more difficult to find. Thus we prefer to use a single array.

\subsection{Plaquette}

\begin{figure}
\center
    \includegraphics[width=8cm]{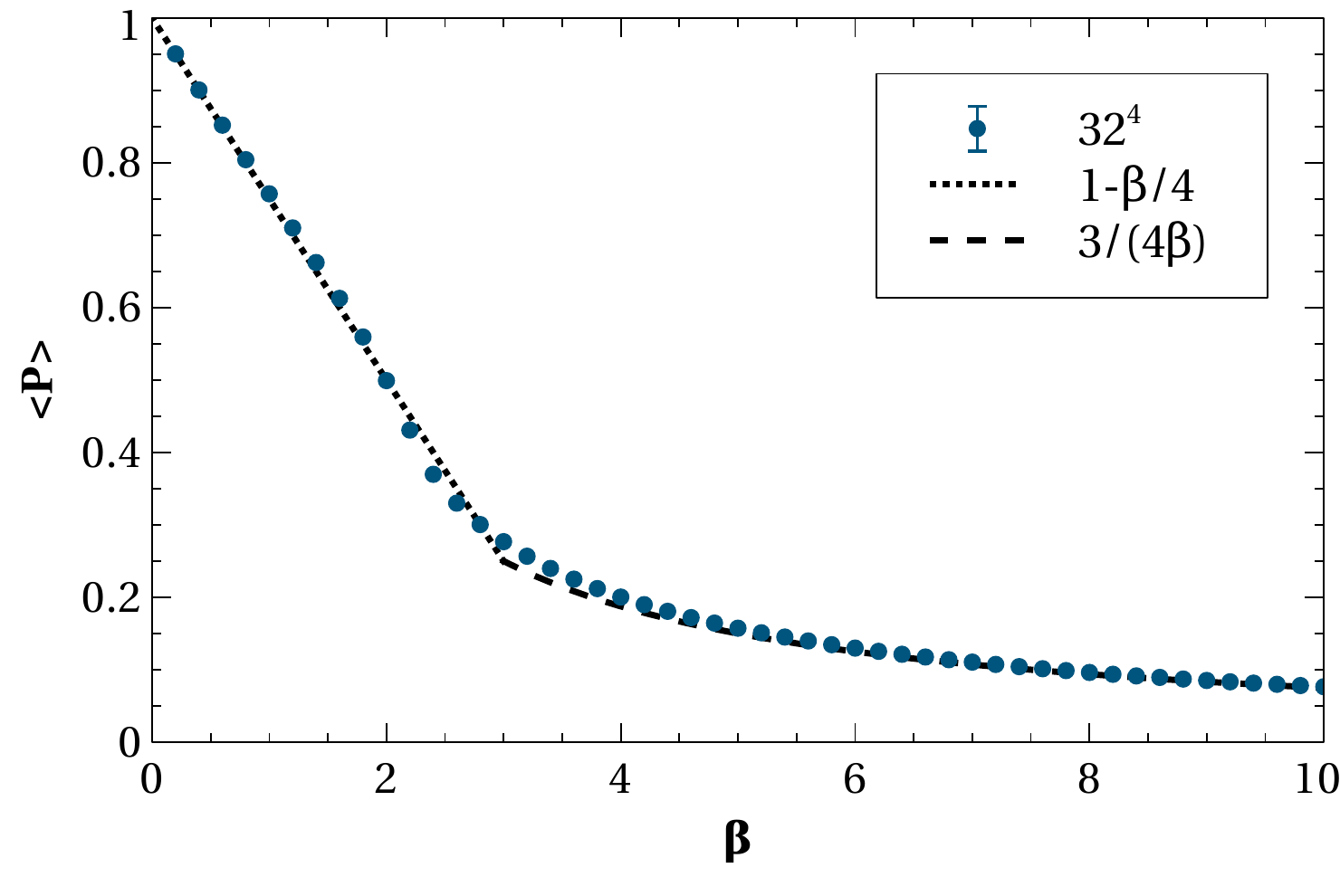}
    \caption{Mean average plaquette for $32^4$ lattice size (data points) and analytic predictions (denoted by dashed lines).}
    \label{fig:mean_avg_plaq}
\end{figure}

The measurement of the average plaquette is defined as the average trace of each plaquette, as defined in Eq. (\ref{eq:plaq}), in all configurations and is the simplest measurement that can be done in the lattice. In Fig. \ref{fig:mean_avg_plaq}, we present the results for the mean average plaquette, as well as the analytic predictions, for different $\beta$, with $10\ 000$ configurations and $32^4$ lattice size.

We are able to perform, at least 3 million Monte Carlo steps per day and calculate the mean average plaquette, in the case of a $32^4$ lattice using the two Fermi GPUs. For a $64^4$ lattice size, we perform $250\ 000$ iterations per day.

\subsection{Polyakov Loop}

\begin{figure}
\center
    \includegraphics[width=8cm]{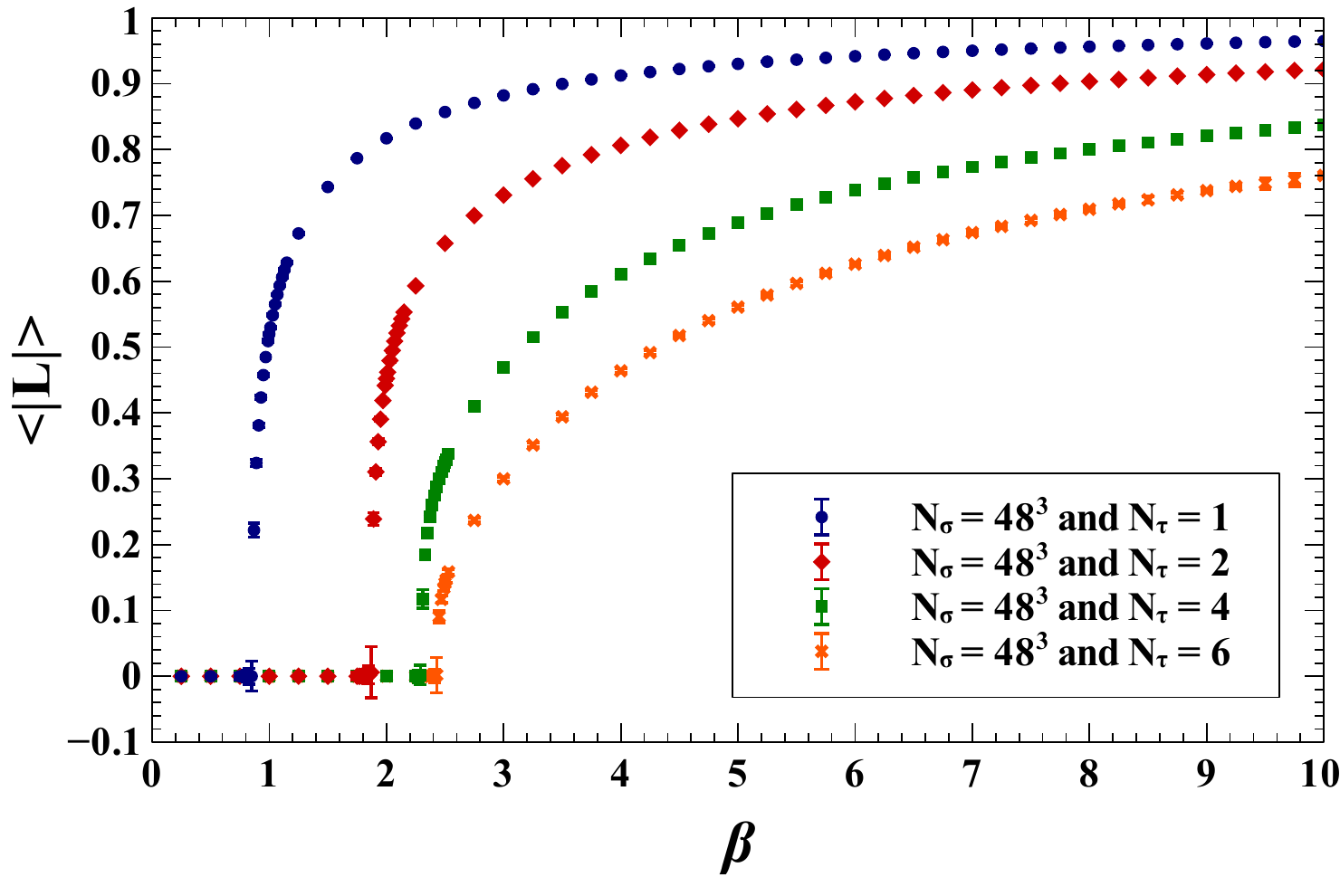}
    \caption{$\beta$ dependence of the mean average Polyakov loop from Monte Carlo simulation.}
    \label{fig:polyakov_loop}
\end{figure}

We now test the GPU performance measuring the Polyakov loop at each generated lattice SU(2), in the same conditions made in the performance tests of Subsection 5.2 . The performance is almost the same, $1.1\times$, compared with only measuring the average plaquette.
Fig. \ref{fig:polyakov_loop} shows the expectation value of the Polyakov loop as a function of $\beta=4/g_0^2$ ($g_0$ is the coupling constant), for several lattice sizes and using $10\ 000$ configurations. The confinement is evident at high couplings, while the deconfinement occurs at small couplings, i.e., the Polyakov loop is zero at high couplings and then at certain critical coupling value it rises to a finite value.
As can be seen, the shape of the curve depends on the temporal size, related to the temperature $T$ of the lattice,
when the spatial size is kept fixed at $N_\sigma=48^3$.

\subsection{The static quark-antiquark potential}

The static quark-antiquark potential, i. e. the potential between two infinitely heavy quarks, has the following long distance expansion,
\begin{equation}
a\,V(a\,R)= A\,a + \frac{B}{R}+\sigma\, a^2\, R\ ,
\end{equation}
where $V(R)$ is the static quark-antiquark potential, $a$ is the lattice spacing, $A$ is a constant term, $B$ is the coefficient to the Coulomb term, $\sigma$ is the string tension and $R$ is the distance in lattice units.
The extraction of the signal of the static quark potential from thermalized lattice gauge configurations is given by, the effective mass plot,
\begin{equation}
V(R)=ln\frac{\Braket{W(R,T)}}{\Braket{W(R,T+1)}}\ ,
\label{eq:pot_lat}
\end{equation}
since
\begin{equation}
\Braket{W(R,T)}=e^{-T\,V(R)}\ .
\end{equation}

\begin{figure}
\begin{centering}
    \includegraphics[width=9cm]{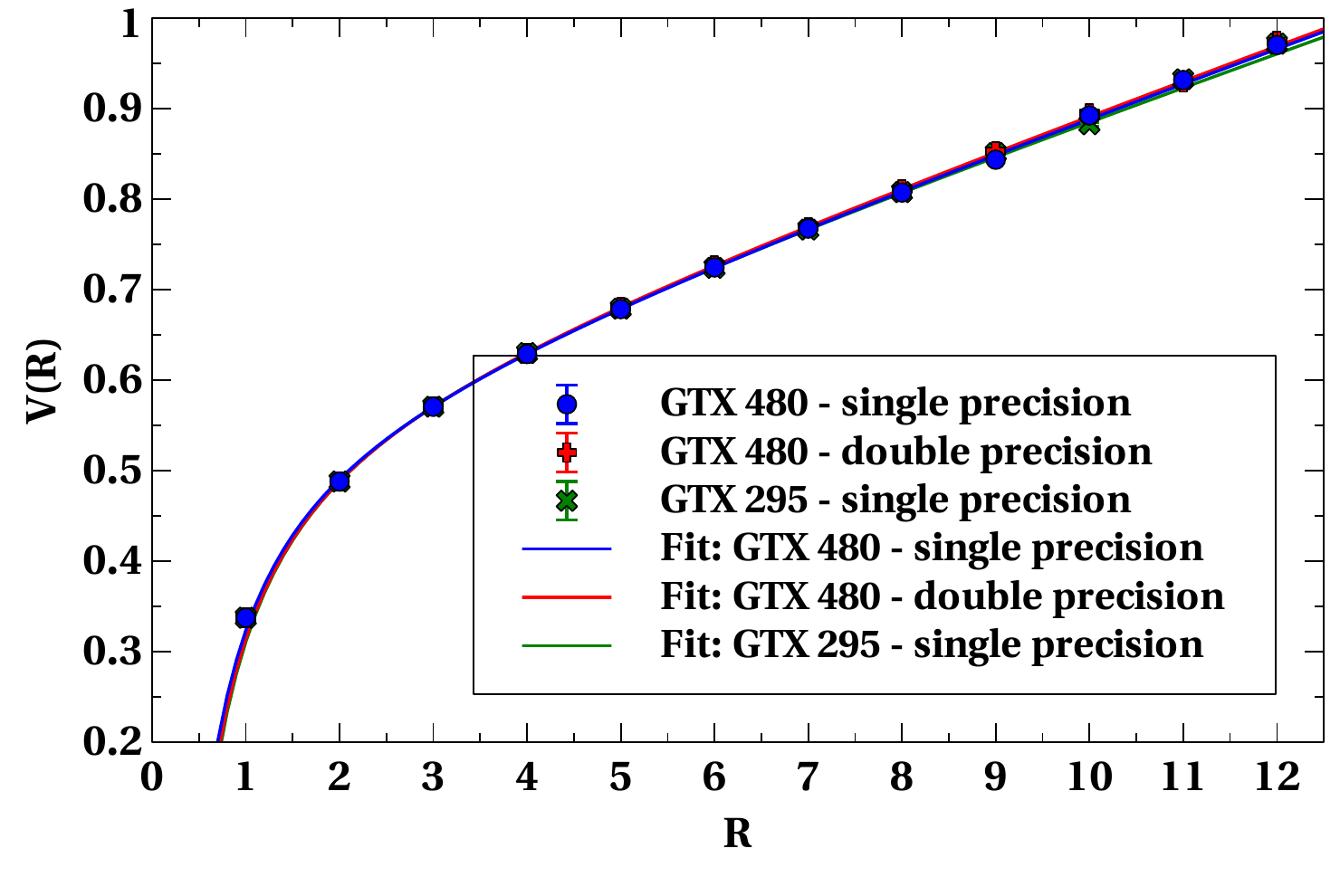}
\par\end{centering}
    \caption{Fit to the static quark-antiquark potential (in lattice units) for 1981 $24^3\times 32$ configurations with $\beta=2.5$ and with APE smearing. Comparison between two different architectures (GT200 and Fermi) in single precision and in double precision for the Fermi architecture.}
    \label{fig:WL_2.5}
\end{figure}

In Fig. \ref{fig:WL_2.5}, we show the fit results for the static quark-antiquark potential using two GPU architectures (GT200 and Fermi). Results in single precision from both architectures are presented, as well as the results from double precison from Fermi architecture. All these results agree within our error bars.

\begin{figure}
\begin{centering}
    \includegraphics[width=9cm]{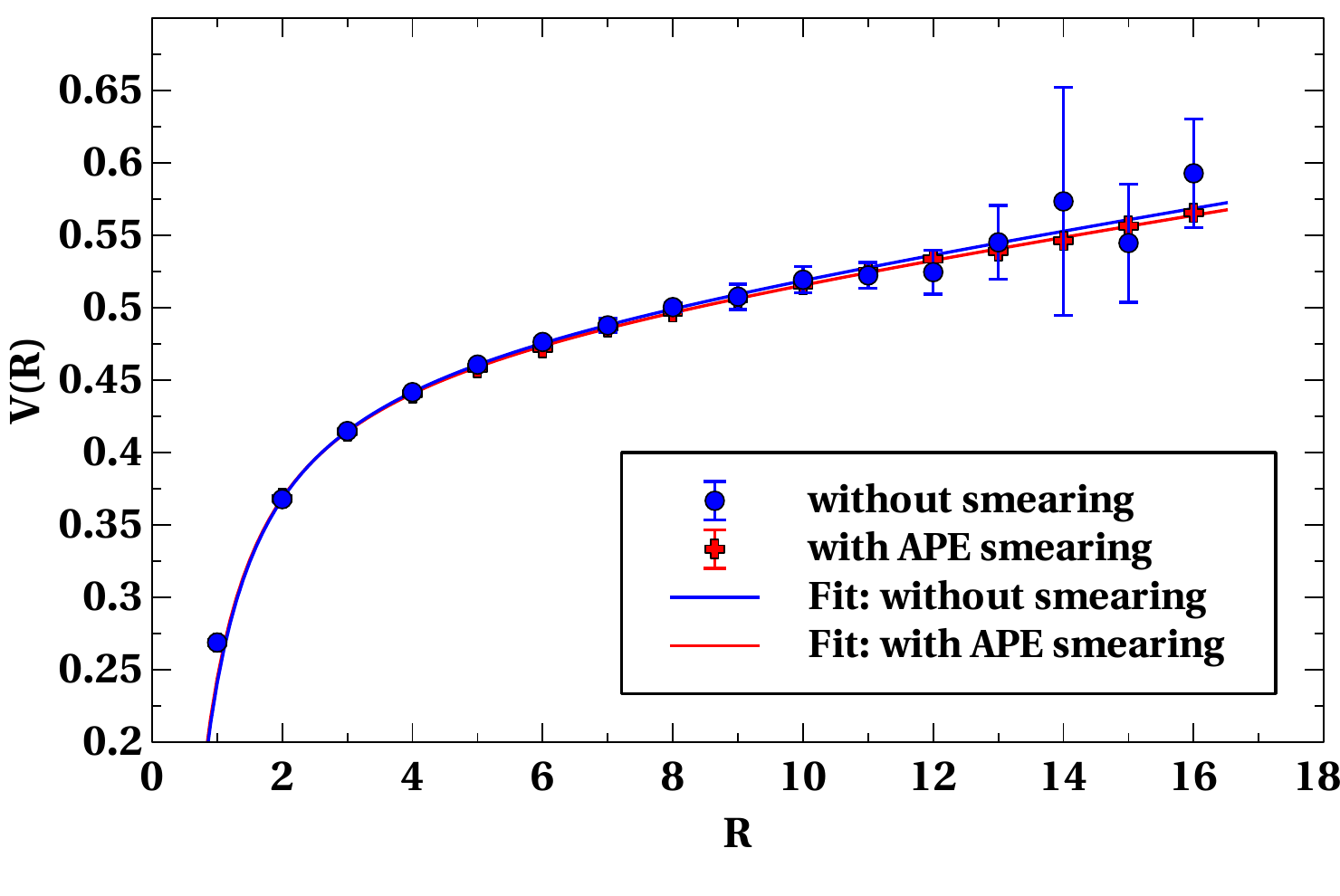}
\par\end{centering}
    \caption{Fit to the static quark-antiquark potential (in lattice units), with and without applying smearing with $\beta = 2.8$ and $24^3\times 48$.}
    \label{fig:WL_2.8}
\end{figure}

In Fig. \ref{fig:WL_2.8}, we show the results for the static quark-antiquark potential with $\beta=2.8$ and $24^3\times 48$  lattice size, using the Fermi GPU. Importantly, we show our results obtained with APE smearing, and without no smearing at all.
In Table \ref{tab:wilson_loop}, we show the values obtained for the lattice spacing $a$ as well as the number of configurations used.
The lattice spacing, $a$, was calculated using the relation $C = \sigma\ a^2$, where $C$ is the value obtained from the linear part of the fit and $\sigma$ the physical value for the string tension, $\sqrt{\sigma}=440\,\text{MeV}$, i.e.,
\begin{equation}
a = \sqrt{C} \frac{197\,\text{MeV}}{440\,\text{MeV}}\ (\text{fm})\ .
\end{equation}

Importantly, utilizing the computational power of the GPUs, we can now afford to calculate the static quark-antiquark potential using thousands of configurations, to study whether the results obtained with and without smearing are in agreement.
Note that the quark-antiquark potential has already been extensively studied \cite{Bhanot:1980fx,Kovacs:1982kz,Stack:1982wb,Huntley1986123,Shakespeare:1998uu},
either for small interquark distances or using different smearing techniques, like the APE smearing,
but usually fail to pick up a significant signal for long distances with no smearing.
The APE smearing, or other smearing method, have the property to enhance the ground state and therefore decouple it from excitations effectively, since the ground state wave function is always the smoothest wave function within any given channel.
The use of APE smearing is an important tool in order to obtain a clear plateau in Eq. (\ref{eq:pot_lat}).

In Table \ref{tab:wilson_loop} for $\beta=2.8$ and in Fig. \ref{fig:WL_2.8} we compare our results with and without smearing.
Although, the unsmeared configurations have larger contribution from the excited states, we can extract the static potential,
noting that the number of configurations needed to obtain a good signal are indeed quite large. We confirm that smearing,
or at least APE smearing, get a potential consistent within error bars to the one produced by unsmeared configurations.

\begin{table}
\begin{centering}
\begin{tabular}{|c|c|c|c|c|c|}
\hline
\T\B$\beta$ & $\sigma a^{2}$ & $a$ (fm) & Lattice size & APE Smearing & \# of config.\tabularnewline
\hline
\hline
\T\B2.5 & $0.036623(625)$ & $0.085682(731)$ & $24^{3}\times32$ & $w=0.2$, $n=25$ & 1981\tabularnewline
\hline
\T\B2.8 & $0.006805(313)$ & $0.036933(850)$ & $24^{3}\times48$ & none & 52712\tabularnewline
\hline
\T\B2.8 & $0.006564(75)$ & $0.036275(207)$ & $24^{3}\times48$ & $w=0.2$, $n=25$ & 1981\tabularnewline
\hline
\end{tabular}
\par\end{centering}
\caption{Lattice spacing results.}
\label{tab:wilson_loop}
\end{table}

\section{Conclusion}

The use of GPUs can improve dramatically the speed of pure gauge SU(2) lattice computations. Using 2 NVIDIA Geforce 480 GTX GPUs in a desktop computer, we achieve $200\times$ the computation speed over one CPU core, in single precision, around 110 Gflops/s using two Fermi GPUs. We obtain excellent benchmarks because our computation is integrally performed in the GPU. Our code can be downloaded from the site of the Portuguese Lattice QCD collaboration \cite{ptqcd}.

The use of textures can increase the speed of memory access when memory access patterns are very complicated and the shared memory cannot be used, although the maximum array size, when using textures, is limited.
Taking advantage of the cache hierarchy introduced in the last architecture, allowed to have similar performance results when accessing to the memory and without having limitations in the array size.

When using multiple GPUs we can improve the speed, making the overlap between computation and data transfers, however this was not yet implemented in the code.
In the future, we will implement this using \mbox{\texttt{cudaMemcpyAsync()}} and streams.
We have used \mbox{\texttt{cudaMemcpy()}} to perform the data transfers. When this function is used, the control is returned to the host thread only after the data transfer is complete.
With \mbox{\texttt{cudaMemcpyAsync()}}, the control is returned immediately to the host thread.
The asynchronous transfer version requires pinned host memory and an additional argument, a stream ID.
A stream is simply a sequence of sorted in time operations, performed in order on the GPU.
Therefore, operations in different streams can be interleaved and in some cases overlapped, a property that can be used to hide data transfers between the host (CPU) and the device (GPU).

We exploit our computational power to compute benchmarks for the Monte Carlo generation of SU(2) lattice gauge configurations, for the plaquette and Polyakov loop expectation values, and for the static quark-antiquark potential with Wilson loops. We are able to verify, utilizing a very large number of configurations, that the APE smearing does not distort the static quark-antiquark potential.

\section*{Acknowledgments}
This work was financed by the FCT contracts POCI/FP/81933/2007, CERN/FP/83582/2008, PTDC/FIS/100968/2008 and CERN/FP/109327/2009.
We thank Marco Cardoso and Orlando Oliveira for useful discussions.


\end{document}